\RequirePackage{fix-cm}
\def\BibTeX{{\rm B\kern-.05em{\sc i\kern-.025em b}\kern-.08em
    T\kern-.1667em\lower.7ex\hbox{E}\kern-.125emX}}
\documentclass[10pt,conference]{IEEEtran}

\usepackage{balance}
\usepackage{graphicx}
\usepackage{textcomp}
\usepackage{color, colortbl}
\usepackage{paralist}
\usepackage{enumitem}
\usepackage{soul}
\usepackage{tcolorbox}
\usepackage{booktabs, multirow}
\usepackage{framed}
\usepackage{xspace}
\usepackage{subfigure}
\usepackage{adjustbox}
\usepackage{url}
\usepackage{lscape}
\usepackage{ifthen}
\usepackage{paralist}
\usepackage{changepage}
\usepackage[super]{nth}
\usepackage[square,numbers]{natbib}
\usepackage{tikz}
\usepackage{todonotes}
\usepackage{threeparttable} 
\usepackage[numbers]{natbib}
\usepackage{pifont}
\usepackage{xcolor} 
\usepackage{fontawesome}  

\usepackage{hyperref}
\hypersetup{
     colorlinks   = true,
     citecolor    = black,
     linkcolor    = blue,
     urlcolor=blue
}

\usepackage{cleveref}

\newcommand{\ie}{\textit{i.e.},\xspace}
\newcommand{\eg}{\textit{e.g.},\xspace}
\newcommand{\etc}{\textit{etc.}\xspace}

\newcommand\pquote[1]{``{#1}''\xspace}
\newcommand{\yes}{\ding{52}}
\newcommand{\no}{\ding{54}}

\definecolor{gray50}{gray}{.5}
\definecolor{gray40}{gray}{.6}
\definecolor{gray30}{gray}{.7}
\definecolor{gray20}{gray}{.8}
\definecolor{gray10}{gray}{.9}
\definecolor{gray05}{gray}{.95}

\newlength\Linewidth
\def\findlength{\setlength\Linewidth\linewidth
	\addtolength\Linewidth{-4\fboxrule}
	\addtolength\Linewidth{-3\fboxsep}
}

\definecolor{backg}{RGB}{225,236,244}
\definecolor{tagtxt}{RGB}{0,0,0}
\definecolor{whitetagtxt}{RGB}{255,255,255}
\definecolor{contextBuilding}{RGB}{238,174,15}
\definecolor{codeInspection}{RGB}{112,33,12}
\definecolor{discussionManagement}{RGB}{103,36,176}
\definecolor{codeReading}{RGB}{43,64,128}
\definecolor{linear}{RGB}{154,201,123}
\definecolor{difficultyBased}{RGB}{124,151,211}
\definecolor{chunking}{RGB}{27,137,120}
\definecolor{testing}{RGB}{252,106,152}
\definecolor{decision}{RGB}{53,65,87}
\definecolor{border}{RGB}{0,0,0}
\definecolor{circleborder}{RGB}{166,0,35}

\newcommand*\tagI{\tikz[baseline=(char.base)]{
            \node[shape=rectangle,text=tagtxt, fill=contextBuilding, draw=border,inner sep=2pt,rounded corners=1mm,] (char) {Context Building};}}

\newcommand*\tagII{\tikz[baseline=(char.base)]{
            \node[shape=rectangle,text=whitetagtxt, fill=codeInspection, draw=border,inner sep=2pt,rounded corners=1mm,] (char) {Code Inspection};}}

\newcommand*\tagIII{\tikz[baseline=(char.base)]{
            \node[shape=rectangle,text=whitetagtxt, fill=decision, draw=border,inner sep=2pt,rounded corners=1mm,] (char) {Decision};}}

\newcommand*\tagIIDM{\tikz[baseline=(char.base)]{
            \node[shape=rectangle,text=whitetagtxt, fill=discussionManagement, draw=border,inner sep=2pt,rounded corners=1mm,] (char) {Discussion Management};}}

\newcommand*\tagIICR{\tikz[baseline=(char.base)]{
            \node[shape=rectangle,text=whitetagtxt, fill=codeReading, draw=border,inner sep=2pt,rounded corners=1mm,] (char) {Code Reading};}}

\newcommand*\tagIICRL{\tikz[baseline=(char.base)]{
            \node[shape=rectangle,text=tagtxt, fill=linear, draw=border,inner sep=2pt,rounded corners=1mm,] (char) {Linear};}}

\newcommand*\tagIICRD{\tikz[baseline=(char.base)]{
            \node[shape=rectangle,text=tagtxt, fill=difficultyBased, draw=border,inner sep=2pt,rounded corners=1mm,] (char) {Difficulty-based};}}

\newcommand*\tagIICRC{\tikz[baseline=(char.base)]{
            \node[shape=rectangle,text=whitetagtxt, fill=chunking, draw=border,inner sep=2pt,rounded corners=1mm,] (char) {Chunking};}}

\newcommand*\tagIIT{\tikz[baseline=(char.base)]{
            \node[shape=rectangle,text=tagtxt, fill=testing, draw=border,inner sep=2pt,rounded corners=1mm,] (char) {Testing};}}

\newcommand*\circled[1]{\tikz[baseline=(char.base)]{
            \node[shape=rectangle,draw=circleborder,inner sep=2pt,rounded corners=1mm,line width=0.5mm] (char) {#1};}}

 \newboolean{journal_version}
 \setboolean{journal_version}{false} 
 \ifthenelse{\boolean{journal_version}}
   {
    \newcommand{\journal}[1]{#1}
    }
   {
    \newcommand{\journal}[1]{}
    }

\newboolean{highlight}

\setboolean{highlight}{false} 

\ifthenelse{\boolean{highlight}}
  {
   \newcommand{\rev}[1]{\textcolor{red}{#1}}
   \newcommand\remove[1]{\textcolor{gray}{#1}}
   }
  {
   \newcommand{\rev}[1]{#1}
   \newcommand\remove[1]{}}

\newcommand{\rqOne}{How do reviewers scope code review comprehension?}
\newcommand{\rqTwo}{What strategies do developers use to perform code review?}
\newcommand{\rqThree}{What are the roles of information sources, knowledge base, and mental models in the code review process?}

\begin{document}

\title{Code Review Comprehension: Reviewing Strategies Seen Through Code Comprehension Theories}



\author{%
\IEEEauthorblockN{Pavl\'{i}na Wurzel Gon\c{c}alves}
\IEEEauthorblockA{\textit{Department of Informatics} \\
\textit{University of Zurich}\\
Zurich, Switzerland \\
p.goncalves@ifi.uzh.ch
}
\and
\IEEEauthorblockN{Pooja Rani}
\IEEEauthorblockA{\textit{Department of Informatics} \\
\textit{University of Zurich}\\
Zurich, Switzerland \\
rani@ifi.uzh.ch
}
\and
\IEEEauthorblockN{Margaret-Anne Storey}
\IEEEauthorblockA{\textit{Department of Computer Science} \\
\textit{University of Victoria}\\
Victoria, Canada \\
mstorey@uvic.ca
}
\and
\IEEEauthorblockN{Diomidis Spinellis}
\IEEEauthorblockA{\textit{Department of Management Science and Economy} \\
\textit{Athens University of Economics and Business}\\
Athens, Greece \\
dds@aueb.gr
}
\and
\IEEEauthorblockN{Alberto Bacchelli}
\IEEEauthorblockA{\textit{Department of Informatics} \\
\textit{University of Zurich}\\
Zurich, Switzerland \\
bacchelli@ifi.uzh.ch
}}

\maketitle

\begin{abstract}
Despite the popularity and importance of modern code review, the understanding of the cognitive processes that enable reviewers to analyze code and provide meaningful feedback is lacking. 
To address this gap, we observed and interviewed ten experienced reviewers while they performed 25 code reviews from their review queue.
Since comprehending code changes is essential to perform code review and the primary challenge for reviewers, we focused our analysis on this cognitive process.
Using Letovsky's model of code comprehension, we performed a theory-driven thematic analysis to investigate how reviewers apply code comprehension to navigate changes and provide feedback. 

Our findings confirm that code comprehension is fundamental to code review. We extend Letovsky's model to propose the Code Review Comprehension Model and demonstrate that code review, like code comprehension, relies on opportunistic strategies. These strategies typically begin with a context-building phase, followed by code inspection involving code reading, testing, and discussion management. To interpret and evaluate the proposed change, reviewers construct a mental model of the change as an extension of their understanding of the overall software system and contrast mental representations of expected and ideal solutions against the actual implementation.
Based on our findings, we discuss how review tools and practices can better support reviewers in employing their strategies and in forming understanding.\\
\emph{Data and material:} \url{https://doi.org/10.5281/zenodo.14748996}
\end{abstract}


\section{Introduction}
\label{sec:intro}
Modern code review is a widely used practice to ensure the quality of code contributions~\cite{sadowski2018modern, gousios2014exploratory, davila2021systematic}.
Typically, developers manually review code changes through an informal, tool-based, and asynchronous process~\cite{gousios2014exploratory, sadowski2018modern}.
Reviewing code changes is seen as beneficial to identify defects and find alternative solutions, improve transparency in the team, and share knowledge among developers~\cite{bacchelli2013expectations}. 
Despite its benefits, code review is time-consuming~\cite{bacchelli2013expectations}, challenging to adopt~\cite{baum2016factors}, and costly to sustain~\cite{oram2010making}. Therefore, researchers have been focused on investigating ways to support developers in performing code reviews efficiently and effectively.

Code comprehension~\cite{soloway1984empirical, heinonen2023synthesizing}, particularly comparative code comprehension~\cite{middleton2022understanding}, is the most important competency reviewers need~\cite{wurzel2023competencies} and simultaneously their greatest challenge~\cite{bacchelli2013expectations}. However, there is little insight into how developers form their understanding during code review and use it to provide feedback.
Understanding these aspects of individual code review performance can inform the design of code review tools and practices~\cite{winograd1997challenge} that support reviewers' construction of the mental models~\cite{storey1999cognitive} and in following effective individual reviewing strategies~\cite{latoza2020explicit}. Human-centric design can improve developers' productivity in code review and their experience when facing common challenges, like code change complexity~\cite{kononenko2016code}.

In this study, we investigate in detail how developers form and use their understanding of the code change under review.
Previous work has studied the behavior of reviewers using various methods, ranging from controlled experiments~\cite{gonccalves2022explicit, fregnan2022first, dunsmore2000role}, to eye-tracking~\cite{uwano2006analyzing}, and analyses of traces left in software repositories~\cite{braz2022less, pascarella2018information, fregnan2022first}.
\rev{These studies describe reviewing mostly as linear reading of code~\cite{fregnan2022first, uwano2006analyzing}. However, the strategy for reviewing large changes remains unexplained (\eg in terms of reading order~\cite{baum2017optimal}).}
\rev{These studies often involve participants reviewing small code changes from unknown systems and lack a real context, purpose, and interactions among authors and reviewers.}\remove{: eye-tracking lacks real review context, and traces in software repositories do not include the real-time decision points and the rationale behind navigation choices among information sources and through the code change.}
As an alternative, we \rev{selected observation accompanied by think-aloud protocols and interviews to capture the reviewer's decisions or comprehension strategies in real time~\cite{aniche2021developers,von1993program}.} \rev{We observed ten experienced developers---recommended by others as great reviewers---while performing 25 code reviews in their work environment to gain insight into their reviewing strategies.}
\remove{The observations were complemented by a think-aloud protocol and a follow-up interview.} We performed an in-depth theory-driven qualitative analysis to code and interpret the collected data, using several theories from code comprehension and psychology~\cite{letovsky1987cognitive, so1964cognitive, rajlich2002role, bak2014self}.

As a result of our observations, we contribute with an extension of Letovsky's code comprehension model~\cite{letovsky1987cognitive}: a \emph{Code Review Comprehension Model}. Our model describes and enhances our understanding of how reviews are performed and of the role of code comprehension in shaping the review process.
The model highlights the opportunistic nature of code review comprehension strategies employed by reviewers to deal with review complexity, \eg scoping down the review or employing other code inspection strategies apart from linear reading, such as chunking or segmenting code based on reviewing difficulty.
We also identify the knowledge and information sources that help reviewers navigate code, construct a mental model of the pull requests (PRs), and provide feedback by comparing their mental model of the PR to ideal and expected solutions.
Finally, based on our observations and analyses, we propose guidelines for performing code review and designing better human-centered review tools.

\section{Background}
\label{sec:background}
We provide details on theories and findings we used in the qualitative analysis to code and interpret the data. 

\smallskip
\noindent\textbf{Opportunistic strategies to code comprehension.}
Code comprehension research has observed, described, and interpreted code navigation strategies aimed at understanding a code artifact~\cite{brooks1977towards, cherubini2007let, corbi1989program, heinonen2023synthesizing}.
Accordingly, researchers have proposed numerous comprehension models to capture the comprehension process~\cite{brooks1977towards, soloway1984empirical, letovsky1987cognitive, von1993program}.
For instance, researchers found that developers initially take a top-down comprehension approach to gain an overall big picture and then move to detailed code snippet~\cite{brooks1977towards}. The top-down approach relies on applying domain knowledge, knowledge of programming plans, and rules of programming discourse to interpret specific implementations of programmatic solutions.
Developers also use other strategies for code navigation, for example, bottom-up~\cite{von1995program, soloway1984empirical} or following the control flow~\cite{pennington1987stimulus}.

Developers \textit{opportunistically} choose the strategies to combine the top-down and bottom-up processes and seek the information relevant to their task to form understanding most efficiently \rev{and update their knowledge through the comprehension process}~\cite{lawrance2010programmers, letovsky1987cognitive}.
In addition to understanding code, code comprehension is a fundamental step in supporting subsequent software development activities~\cite{von1995program}. Relevant to our study, code comprehension also determines how developers review code, the review outcomes, and the artifacts created in the process. We aim to understand which approaches are used and how the mental model of the reviewed code change is formed.

\smallskip
\noindent\textbf{Role of experience in code review comprehension.} Code review is based on forming an accurate mental model of the change and ensuring maintainability by other developers~\cite{wurzel2023competencies}.
Developers achieve understanding through recognition of programming plans, \ie stereotypical implementations of goals~\cite{soloway1984empirical}.
Reviewers' experience is crucial to recognize and correct programming patterns into readable and maintainable code: Novice developers tend to have more difficulty understanding a program~\cite{boehm1992role}, gaining the overall picture, and identifying programming patterns~\cite{jeffries1982comparison}.
Consequently, adherence to coding conventions and programming patterns ensures also adherence to representations shared throughout the developer community and thus, fitting the mental schemas of other developers.
In the analysis, we have viewed reviewers as gatekeepers ensuring code understandability and maintainability by promoting adherence to these shared standards.

\smallskip
\noindent\textbf{Code Review Strategies.} Past research has provided evidence that reviewers often perform \textit{ad hoc} reviews~\cite{ciolkowski2003software}: They ``just read the code'' or adopt an unsystematic approach that relies on personal strategies to navigate code and identify defects~\cite{ciolkowski2003software, uwano2006analyzing}.
\citet{baum2017optimal} found that reviewers mostly navigate the files linearly following the order offered by the review tool. \citet{fregnan2022first} found a linear decline in the number of comments reviewers leave as the files appear later in a change set, and measured diminished developers' effectiveness at detecting defects in files presented last by review tools.
In eye-tracking experiments, reviewers were found to perform reviews mostly linearly, splitting the review into a scanning phase (where reviewers first get an overview of the code) and a detailed phase (where they return to look into specific code segments)~\cite{uwano2006analyzing}.

Although review navigation seems to be often linear, this may not be the case for more complex reviews. In fact, complex changes are more challenging to review~\cite{kononenko2016code} and the order in which reviewers choose to review them remains largely unexplained~\cite{baum2017optimal}. For example, \citet{spadini2019test} indicate that reviewers might choose to follow diversified review strategies, such as performing test-driven reviews.
Therefore, in our study we observe reviews of changes of varying complexity to better understand how complexity may affect the process.

\smallskip
\noindent\textbf{Code comprehension and code review.}
Similarly to code comprehension, it seems reasonable to expect that reviewers performing ad hoc reviews~\cite{ciolkowski2003software} may shape their reviewing process through a combination of strategies and opportunistic decisions that aim to optimize for both effective understanding and efficient reviewing.
However, code review requires deeper engagement of higher-level cognitive processes (\eg decision making and analysis) than code comprehension alone due to the need to inspect the code changes for quality aspects~\cite{floyd2017decoding}.

Code review is a specific context for using code comprehension, as understanding occurs within an environment where the changes to the software system are \emph{iterative} (reviewers repeatedly comprehend the same artifact), \emph{incremental} (they comprehend a modification of a likely partially known software system), and \emph{interactive} (the comprehension is supported by interactions with the author and other colleagues).
Reviewing may require developers to work with other software artifacts, such as the PR discussion, issues, testing tools, architecture, and design documentation. Therefore, comprehension models that address only code navigation may be insufficient to fully capture code review comprehension.


\section{Methodology}
\label{sec:methods}

\begin{figure}[t]
    \centering
    \includegraphics[width=\linewidth]{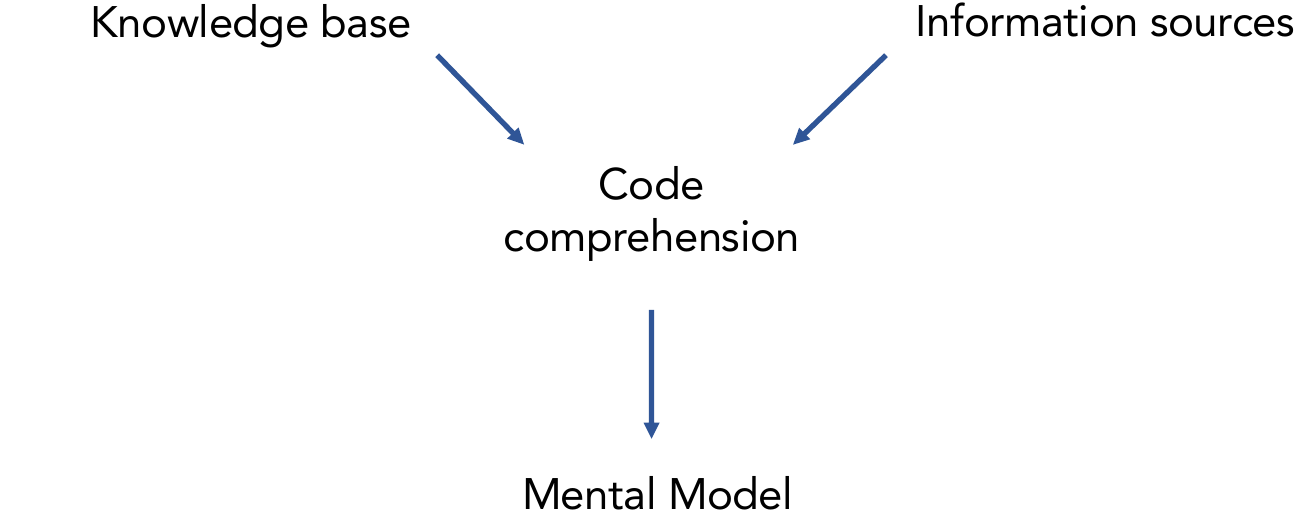}
    \caption{Letovsky's Model of Code Comprehension~\cite{letovsky1987cognitive} - Code comprehension is an assimilation process using knowledge base and information sources to create a mental model of the code.}
    \label{fig:letovsky}
\end{figure}

Our study investigates how software developers perform code reviews through the lens of code comprehension theories. As the main tool for approaching the data and informing the research questions, we use \citeauthor{letovsky1987cognitive}'s model of code comprehension~\cite{letovsky1987cognitive} (shown in \Cref{fig:letovsky}). The model is based on \citeauthor{so1964cognitive}'s model of cognitive development~\cite{so1964cognitive}, which posits that learning is an activity with a \emph{scope}, as the ability to learn in a certain moment has a limit of information that can be effectively processed.

\subsection{Research Questions}

We address three research questions. First, we ask:

\medskip
\noindent\textbf{RQ$_1$. \emph{\rqOne}}


Although reviews are often seen as unsystematic~\cite{ciolkowski2003software, uwano2006analyzing}, their reliance on code comprehension~\cite{bacchelli2013expectations} suggests that reviewing may follow opportunistic strategies as code comprehension does.
Understanding review strategies and the decisions in their selection, particularly when handling complex changes~\cite{baum2017optimal}, can provide valuable insights on how to support reviewers to perform effective code reviews and inform the design of more human-centered review tools. In RQ$_2$, we focus on the Code Comprehension component of \citeauthor{letovsky1987cognitive}'s model:

\medskip
\noindent\textbf{RQ$_2$. \emph{\rqTwo}}
\smallskip


In \citeauthor{letovsky1987cognitive}'s model of code comprehension~\cite{letovsky1987cognitive} (see \Cref{fig:letovsky}), code comprehension is accompanied by other three components of the model: (1) the information sources including the code artifact being understood and (2) the knowledge base, which interact in the code comprehension process to create (3) the mental model of the code artifact.
To capture how code comprehension is used in code reviews, we also ask:

\medskip
\noindent\textbf{RQ$_3$. \emph{\rqThree}}
\smallskip

\subsection{Data Collection}

Given that the cognitive processes underlying a code review are not directly observable, we use observation, interviewing, and think-aloud protocols to capture the reviewer's cognitive process similarly to other studies looking into code comprehension or developers' approaches to perform engineering tasks~\cite{aniche2021developers,von1993program}.
Throughout the paper, participants and specific review sessions are referenced using the format [participant ID][review number] (\eg P2R3).

\paragraph{Recruitment of Participants}
\rev{Following estimates for grounded theory studies, we expected to need 20 to 30 observations to reach saturation~\cite{creswell2015educational}. We used purposeful sampling method~\cite{baltes2022sampling} to achieve high variety in the data by observing 1) reviews in a real-world context, thus observing more varied PRs, 2) reviews of varying complexity and 3) reviews in both OSS community and in in-house software development teams.}

We recruited reviewers recommended as great reviewers or who had at least ten years of code review experience. This approach allowed us to observe experienced reviewers with purposeful decision-making in code reviews who have high knowledge of programming patterns and are recognized by other people as insightful. We specifically obtained six reviewers through their companies---four from an open-source software development company (participants P1, P3, P4, P5) and two from a closed-source development company (P8 and P10). We asked developers within these companies to recommend great reviewers, and researchers only received contact details for those who had been recommended and agreed to participate. Additionally, four reviewers were contacted individually: P2, an experienced author of open-source software (OSS) and of technology books and academic publications on software quality; P6, an experienced OSS reviewer recommended by P2; P7, a software engineer at a big technological firm who shares their coding and reviewing practices on their YouTube channel; and P9, an insightful and experienced reviewer known from a previous study.
The reviewers typically held leadership and mentoring roles in their teams, had over ten years of professional experience, and reviewed code daily (see \Cref{tab:demographics}).

\paragraph{Review\&Interview sessions}
We collected data in three phases -- (1) observation of code review sessions, (2) exploratory interview, and (3) a demographic survey. We conducted each session through an online video call that lasted approximately one hour. With the consent of the participants, we recorded each session for subsequent data analysis. During each session reviewers performed both a short and a long review on PRs from their review queue (at work or in their open source project). While reviewing, participants used a think-aloud method. \rev{The interviewer played a passive role, mainly intervening only to facilitate the think-aloud. After each review, the researcher conducted a short semi-structured interview to clarify the observations -- establishing the sequence of actions/steps in the review, reasoning and choices of the reviewers and differences among participants. These areas were captured in a non-binding interview guide.}
Following these sessions, the participants completed a short demographic survey.

As a result of our data collection process, we acquired rich data from 25 review sessions. 
\Cref{tab:reviews} details the reviewed changes: Their sizes ranged from 2 to 34,520 changed lines of code, and the reviews were conducted using GitHub (N=18), GitLab (N=2), Phabricator(N=2), or Gerrit(N=3). We observed 15 reviews in open-source and 10 in closed-source contexts. In most reviews, our participants ended their task by requesting changes (N=11) or providing comments (N=9), while they directly approved a minority (N=5).

\paragraph{Transcription}
We transcribed the observations and interviews to enable the subsequent analyses. For the observations, we created the transcripts based on what was vocalized by the reviewers in their think-aloud and by actions performed in the reviewing platform. As a validation of this transcription process, the second author of the study compared three transcripts (P2R2, P4R1, P9R1) against the non-anonymized observations, providing feedback to the first author and pointing out additional insights. These were used to improve and update the other transcripts.




\begin{table*}[ht]
\centering
\caption{Descriptive demographics of the study participants}
\label{tab:demographics}
\begin{tabular}{lll|r|r|r|l|l}
\textbf{Participant} & \multirow{2}{*}{\textbf{Role}} & \multirow{2}{*}{\textbf{Gender}} & \multicolumn{2}{c|}{\textbf{Experience (years)}} & \textbf{Team/Project} & \multicolumn{2}{c}{\textbf{Frequency of}} \\
\textbf{ID} & & & \textbf{programming} & \textbf{reviewing} &  \textbf{tenure (years)} & \textbf{programming} & \textbf{reviewing} \\ \midrule
P1                & Technical Lead   & Female          & 17                                                   & 17                                      & 3                            & Daily                          & Daily                          \\
P2                & Code base author & Male            & 37                                                   & 10                                      & 1                            & Weekly                         & Monthly                        \\
P3                & Project Lead     & Male            & 17                                                   & 5                                       & 3                            & Daily                          & Daily                          \\
P4                & Team Lead        & Male            & 16                                                   & 15                                      & 10                           & Daily                          & Daily                          \\
\rowcolor[HTML]{EFEFEF} 
P5                & \multicolumn{7}{c}{\cellcolor[HTML]{EFEFEF}\textit{Chose not to disclose}}                                                                                                                                                                    \\
P6                & Core contributor & Male            & 14                                                   & 16                                      & 13                           & Daily                          & Daily                          \\
\rowcolor[HTML]{EFEFEF} 
P7                & \multicolumn{7}{c}{\cellcolor[HTML]{EFEFEF}\textit{Chose not to disclose}}                                                                                                                                                                    \\
P8                & Senior developer & Male            & 3                                                    & 2                                       & 1                            & Daily                          & Daily                          \\
P9                & Team Lead        & Male            & 11                                                   & 10                                      & 4                            & Daily                          & Daily                          \\
P10               & Team Lead        & Male            & 15                                                   & 9                                       & 2                            & Daily                          & Daily                          \\ \bottomrule
\end{tabular}
\end{table*}

\begin{table*}[ht]
\centering
\caption{Descriptives of the real-world reviews conducted by our participants during the observations.}
\label{tab:reviews}
\begin{tabular}{llllll|r|r|r|r}
\textbf{Review} & \multirow{3}{*}{\textbf{Tool}} & \multirow{3}{*}{\textbf{Code}} & \multirow{3}{*}{\textbf{Iteration(*)}} & \multirow{3}{*}{\textbf{Draft}} & \multirow{3}{*}{\textbf{Verdict}} & \multicolumn{4}{c}{\textbf{Number of}} \\ 
\multirow{2}{*}{\textbf{ID}}& & & & & & \multirow{2}{*}{\textbf{reviewers}} & \multirow{2}{*}{\textbf{files changed}} & \multicolumn{2}{c}{\textbf{lines}} \\ 
& & & & & & & & \textbf{added} & \textbf{removed} \\ \midrule
P1R1               & Gerrit        & CSS           & 2nd       & \no             & Request changes               & 5                           & 3              & 97                 & 86                \\
P1R2               & GitHub        & OSS           & 1st     & \no              & Comment          & 3                           & 1              & 15                 & 21                \\
P1R3               & Gerrit        & CSS           & 1st    & \no              & Request changes               & 4                           & 6              & 174                & 2                 \\ \hline
P2R1               & GitHub        & OSS           & 1st      & \no              & Request changes  & 1                           & 13             & 1,366               & 76                \\
P2R2               & GitHub        & OSS           & 2nd       & \no              & Accept          & 1                           & 8              & 376                & 45                \\
P2R3               & GitHub        & OSS           & 3rd    & \no              & Request changes  & 1                           & 6              & 109                & 39                \\
P2R4               & GitHub        & OSS           & 1st    & \no              & Accept          & 1                           & 1              & 13                 & 4                 \\ \hline
P3R1               & GitHub        & OSS           & 1st    & \no              & Accept          & 1                           & 1              & 4                  & 4                 \\
P3R2               & GitHub        & OSS           & 1st    & \yes            & Request changes  & 1                           & 7              & 369                & 98                \\ \hline
P4R1               & GitHub        & OSS           & 1st    & \no              & Accept          & 1                           & 1              & 1                  & 1                 \\
P4R2               & GitHub        & OSS           & 1st    & \yes            & Comment          & 1                           & 3              & 158                & 92                \\ \hline
P5R1               & Gerrit        & OSS           & 1st   & \no              & Request changes               & 3                           & 12             & 79                 & 0                 \\
P5R2               & GitHub        & OSS           & 2nd       & \yes            & Comment          & 2                           & 36             & 3,848               & 40                \\ \hline
P6R1               & GitLab        & OSS           & 1st    & \yes            & Comment          & 1                           & 4              & 49                 & 6                 \\
P6R2               & GitLab        & OSS           & 1st    & \no              & Comment          & 1                           & 6              & 360                & 3                 \\ \hline
P7R1               & Phabricator   & OSS           & 1st    & \no              & Request changes  & 3                           & 3              & 3                  & 0                 \\
P7R2               & Phabricator   & OSS           & 3rd+    & \no              & Comment          & 6                           & 5              & 57                 & 36                \\ \hline
P8R1               & Github        & CSS           & 3rd+    & \no              & Accept          & 2                           & 1              & 41                 & 0                 \\
P8R2               & Github        & CSS           & 3rd+       & \no              & Request changes  & 3                           & 17             & 535                & 2                 \\ \hline
P9R1               & Github        & CSS           & 1st    & \no              & Comment          & 1                           & 17             & (**)17,030              & (**)17,490             \\
P9R2               & Github        & CSS           & 1st    & \no              & Request changes  & 2                           & 28             & 446                & 165               \\
P9R3               & GitHub        & CSS           & 1st    & \no              & Comment          & 2                           & 56             & 842                & 1,177              \\ \hline
P10R1              & Github        & CSS           & 3rd+    & \no              & Request changes  & 3                           & 19             & 636                & 9                 \\
P10R2              & Github        & CSS           & 2nd    & \no              & Request changes  & 2                           & 1              & 7                  & 5                 \\
P10R3              & Github        & CSS           & 1st    & \no              & Comment          & 3                           & 1              & 1                  & 5 \\
\bottomrule
\end{tabular}
\begin{tablenotes}
\item (*) Iteration number for the participant as a reviewer
\item (**) The code change was extremely large due to the inclusion of a large autogenerated \texttt{.yaml} file.
\end{tablenotes}
\end{table*}

\subsection{Data Analysis}\label{sec:analysis}
Through familiarizing with the collected observations and interviews, we confirmed that themes of understanding and comprehension were prominent and that code comprehension theory, particularly \citeauthor{letovsky1987cognitive}'s model of code comprehension~\cite{letovsky1987cognitive}, provides us with an effective vocabulary to capture the scope of our observations. 

\begin{figure}[t]
    \centering
    \includegraphics[width=\linewidth]{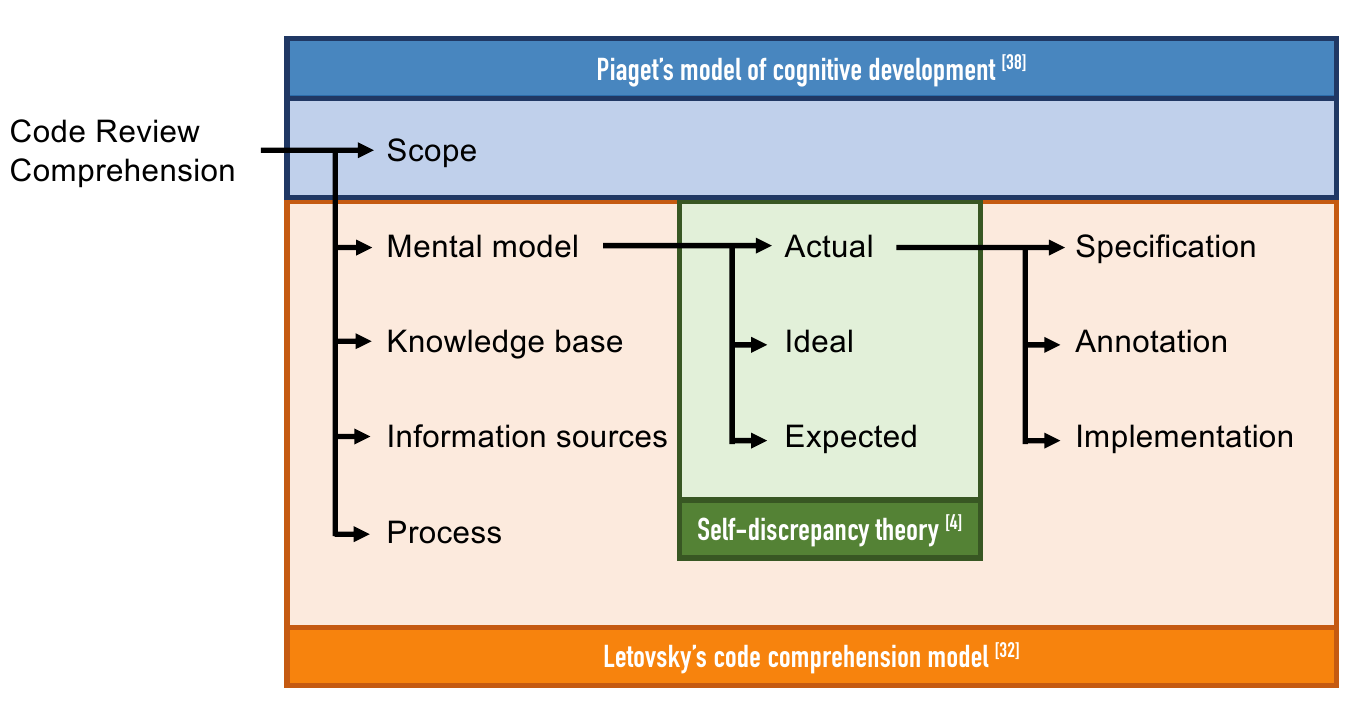}
    \caption{\rev{Coding schema used to structure and code the data according to the selected code comprehension and psychological theories.}}
    \label{fig:schema}
\end{figure}

\rev{Therefore, we used \citeauthor{letovsky1987cognitive}'s model complemented by other models, terms and theories from code comprehension and psychology~\cite{so1964cognitive, bak2014self} to produce the coding schema presented in \Cref{fig:schema}, through which we performed a theory-driven thematic analysis~\cite{braun2006using}.}

\rev{\begin{itemize}[leftmargin=1em]
    \item \emph{\citeauthor{letovsky1987cognitive}'s code comprehension model}~\cite{letovsky1987cognitive} (\Cref{fig:letovsky}): captures code comprehension as an interplay of information sources and knowledge base through which the developer creates a mental model representing the code artifact. Mental models consist of the specification, annotation and implementation layer.
    \item \emph{\citeauthor{so1964cognitive}'s model of cognitive development~\cite{so1964cognitive}:}
    \rev{posits that learning is an activity with a \emph{scope}} that happens in achievable increments rather than being formed as a complete and comprehensive understanding of a new input.
    \item \emph{Self-discrepancy theory~\cite{bak2014self}:} interprets human distress as a result of a discrepancy between their self perception and their own and societal ideals and expectations. We applied it to interpret how reviewers use discrepancies between the PR and their expectations and ideals to interpret and evaluate the code change.
\end{itemize}}

The second author validated the text excerpts using transcripts from three review sessions (P2R2, P4R1, P9R1). Within the concept-related excerpts, we proceeded with a bottom-up coding~\cite{braun2006using} to understand how these concepts can be described and understood in the code review context. Subsequently, we used the research work mentioned in \Cref{sec:background} as a basis to interpret the data.

\rev{The analysis aimed to reach code saturation~\cite{hennink2017code}. Therefore, the themes reported in the paper represent a coding structure that was stable and repetitive by the end of the analysis.}

We performed the analysis using the qualitative research software NVivo 14. The first author was the person collecting and analyzing the data. To support the validity of the findings, we used \emph{peer debriefing}~\cite{creswell2000determining}. Theories and preliminary results were regularly discussed with the second author who also performed checks for validity of transcripts and first-level coding.
We discussed results and interpretations among all the authors at two major milestones: (1) during the analysis and (2) once the findings were interpreted and reported.

\rev{Our replication package includes the observation and interview transcripts, interview guide, coding schema, and other documents~\cite{replication_package}.}

\subsection{Ethics and Data Handling}

The Human Subjects Committee of the home university of the first, second, and last author approved the study design. Reviewers signed an informed consent before participating in a monitored review session.
Observations including participants' screens and faces were temporarily stored on the university server to create anonymized behavioral transcripts.
The observed code came from OSS projects or was shared with the permission of the project/team lead, given no sensitive information was shared in the recorded code.

\subsection{Limitations}
The scope of the study focused on
describing aspects related to code comprehension, using observation of reviews performed by experienced reviewers performed in a real-world context. There are limitations to this focus.

Being the first observational study of cognitive processes in code review, our data collection was initially exploratory 
and focused on aspects of code comprehension only in the analysis phase. We did not delve deeply into
other key aspects of code review, such as reviewers' interactions---a possible rich field for further exploration.

\rev{We conducted the observations via a video call. We asked participants whether their review process was the same as usual. Only P2 mentioned they were highly aware of being observed and possibly made their review more thorough than usual, suggesting a possible influence of the Hawthorne effect~\cite{diaper1990hawthorne}.  To capture natural strategies, we avoided interrupting the reviewers, keeping the clarifications for the end of the sessions. This approach allowed us to observe the higher level flow, rather than to investigate details of each decision made, \eg reasoning for writing individual comments}.

The study participants are experienced reviewers with established reviewing strategies and reasoning. Therefore, this data does not represent the strategies of novice reviewers.

We observed reviews done on four online platforms (GitHub, GitLab, Gerrit, and Phabricator). We did not observe reviews in other contexts, such as IDE-integrated code review tools, reviews through emails, or in-person reviews. Therefore, our observations may not generalize to those contexts.

Since the data was collected and mainly analyzed by the first author, the results are shaped mainly by her knowledge and expertise.
The second author validated the material on two separate occasions for coding and transcription quality. Furthermore, the results and interpretations were repeatedly discussed with all other co-authors in individual and group meetings and updated accordingly.

\section{Results}
\label{sec:results}
The 25 reviews performed by ten expert reviewers led to a total of 14 hours and 42 minutes of recorded observations and follow-up interviews. The coding led to the definition of 846 codes captured in 3,792 unique references. The replication package also includes the list of themes from the analysis~\cite{replication_package}.

\begin{figure}
    \centering
    \includegraphics[width=\linewidth]{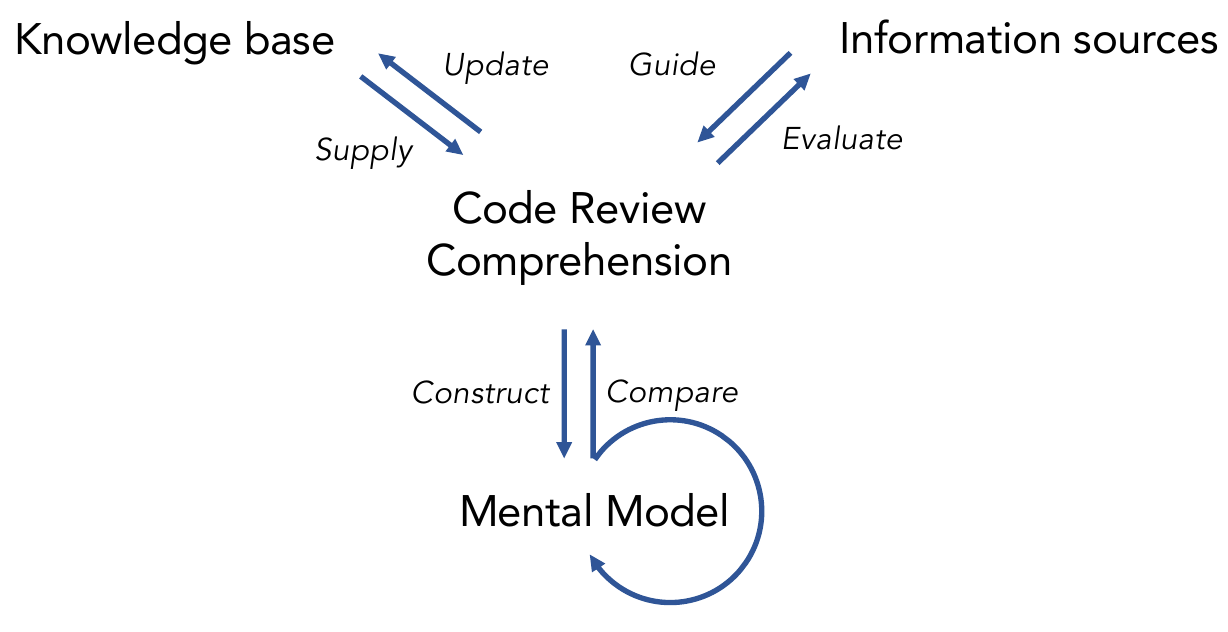}
    \caption{Code Review Comprehension Model - Code Review Comprehension uses opportunistic strategies to enrich information sources through identifying issues, commenting, and proposing improvements.}
    \label{fig:crcm}
\end{figure}

Our analysis showed that the \citeauthor{letovsky1987cognitive}'s code comprehension model~\cite{letovsky1987cognitive} was efficient, yet not complete enough to cover the case of code review comprehension. Therefore, we propose an extended model to fill this gap: the Code Review Comprehension Model (CRCM), as depicted in \Cref{fig:crcm}.
Through our new model, we describe the individual components (\emph{Code Review Process}, \emph{Information Sources}, \emph{Knowledge Base}, and \emph{Mental Model}) alongside their function.
\rev{While Letovsky's model views code comprehension as combining knowledge base and information sources to form a mental model of the code, CRCM also integrates more recent code comprehension theories like constructivism~\cite{rajlich2002role} (\ie comprehension is a learning process that updates the knowledge base) and views code review comprehension as an opportunistic process shaped by the purpose and usage of the understanding~\cite{von1993program} (\ie to understand code changes, select an appropriate reviewing strategy, and provide feedback).}
In the following subsections, we present detailed results for each component of the CRCM, \rev{their interactions,} and the review scope.

\subsection{\textbf{RQ1}: Scoping Code Review Comprehension}
\label{sec:scope}

The goal of RQ1 is to explore the limits reviewers set for their review.
We define review \emph{comprehension scope} as the completeness and depth of the code review comprehension process.
We observed reviewers employing four distinct ways of scoping their review: they performed
\begin{inparaenum}[(1)]
 \item \textbf{full} reviews, \ie the reviewers achieved a complete understanding of the code changes,
 \item reviews \textbf{focused} on specific aspects of the code change (\eg high-level rationale coherence and areas related to their expertise),
 \item \textbf{partial} reviews of sections of the change (sometimes due to the review being terminated prematurely), and
 \item\textbf{shallow} reviews where only a superficial understanding was reached, and reviewers based their judgment on external factors (\eg sufficient testing and the expertise and responsibility areas of the other involved reviewers; as P4 explained: \pquote{If the PR is really really big, I trust in the CI. I trust if all the tests are passed I understand that the changes that are being added are not affecting the current behavior.}).
 \end{inparaenum}

Reviewers reported that several factors---related to the nature of PR-based code review and code complexity---influence the completeness and thoroughness of their reviews.

Given that code review is \emph{iterative} and \emph{incremental} in nature, a full understanding of the code changes can be gradually achieved through multiple review iterations. Reviewers use these characteristics to break up the entire review into manageable steps and to benefit from the expectation of further review iterations. As P5 noted: \pquote{I don't need this knowledge answered right now. I need this to be answered before I approve, but I can live without it today.}
Each review iteration also allowed reviewers to narrow the scope of the review. When reviewing the same PR for the third time (P2R2), P2 explained: \pquote{I can take shortcuts. If [the file] has no comments, I will ignore it, I will not really review it. And this thing converges, right? Because more and more files enter the state.}

Code review is also \emph{interactive}: Reviewers and authors can support each other in understanding the code.
The \emph{collective comprehension} of the author and reviewers can have more importance than an individual reviewer reaching full comprehension of the PR. Reviewers consider the added value their review can provide, the expertise of other reviewers, and the opportunity to seek clarifications from the author. As P9 puts it: \pquote{If this was application code and I didn't understand anything, I would definitely either make sure that I understand or ask about it. Or there is another reviewer who can go a little bit deeper than me.}

The complexity of the change under review fundamentally contributes to the need to scope their review comprehension.
When we asked P5 why they reviewed only a part of the change, they explained: \pquote{I have to ... it is impossible to fit [it] into my head.}
Small changes require less comprehension or reviewer involvement; as P2 put it: \pquote{There are not many things that can go wrong in one line of code and the automated tools would have caught most of the issues already.}

\subsection{\textbf{RQ2}: Strategies Reviewers Use To Perform Code Review}
\label{sec:process}

\emph{RQ$_2$} explores the strategies reviewers use to understand and review the code change.
The code review process, similar to code comprehension, is opportunistic. Reviewers combine various information sources with their knowledge base and existing mental models to understand the code change, evaluate the PR, and provide feedback. They deliberately select their reviewing strategy, accounting for the change's complexity, aligning with their current priorities and review scope.

\begin{figure}[ht]
	\centering
	\subfigure[Activities reviewers combine to reach an understanding and create a reviewing strategy. Each activity is described in the text using its number and colored label.]{\includegraphics[width=\linewidth]{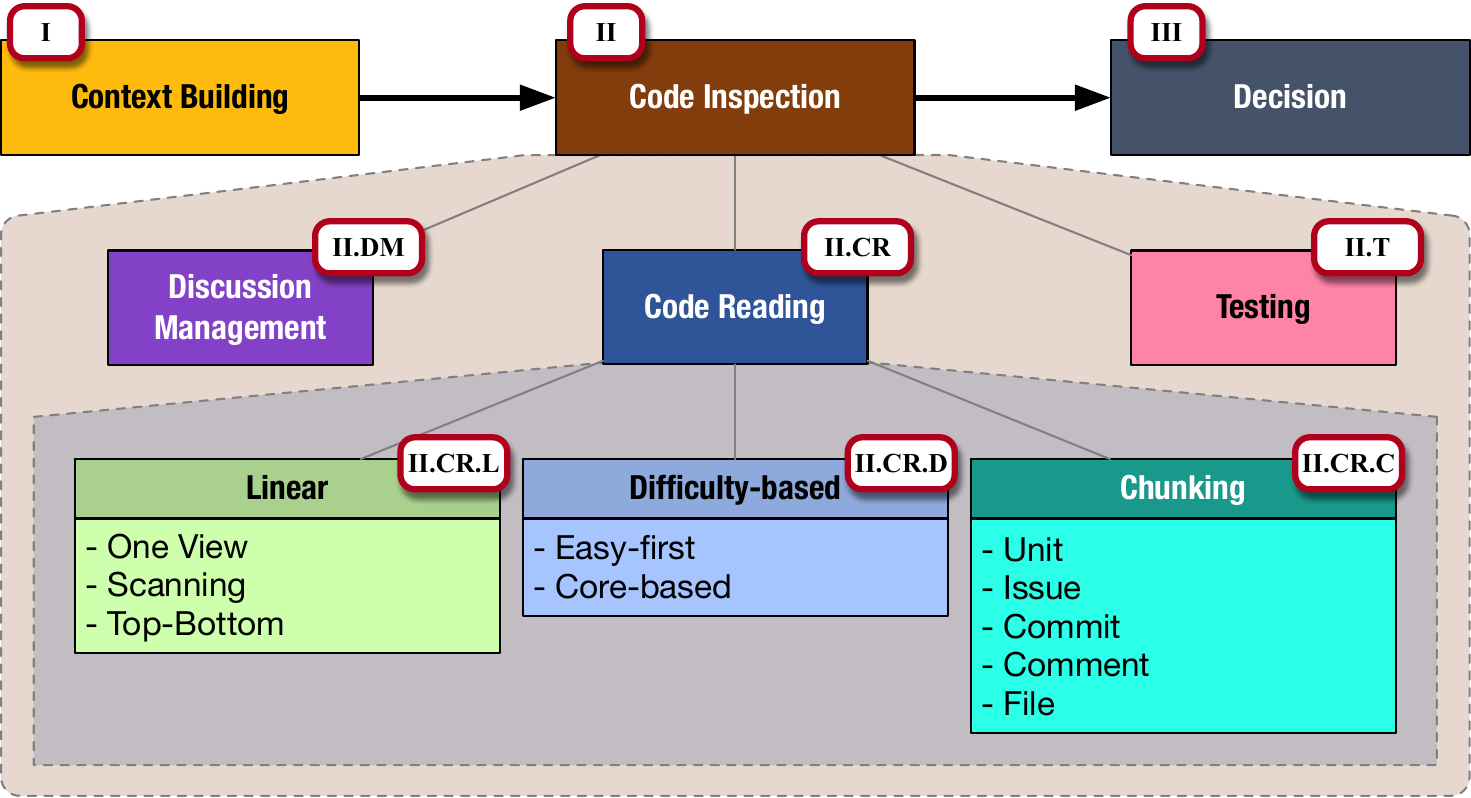}}
    \subfigure[Composition of these activities in each review session. The color legend corresponds to \Cref{fig:process} (a). Some activities can be employed in parallel (represented by split-color fields.) The reviews are ordered by size - smallest to largest code changes.]{\includegraphics[width=\linewidth]{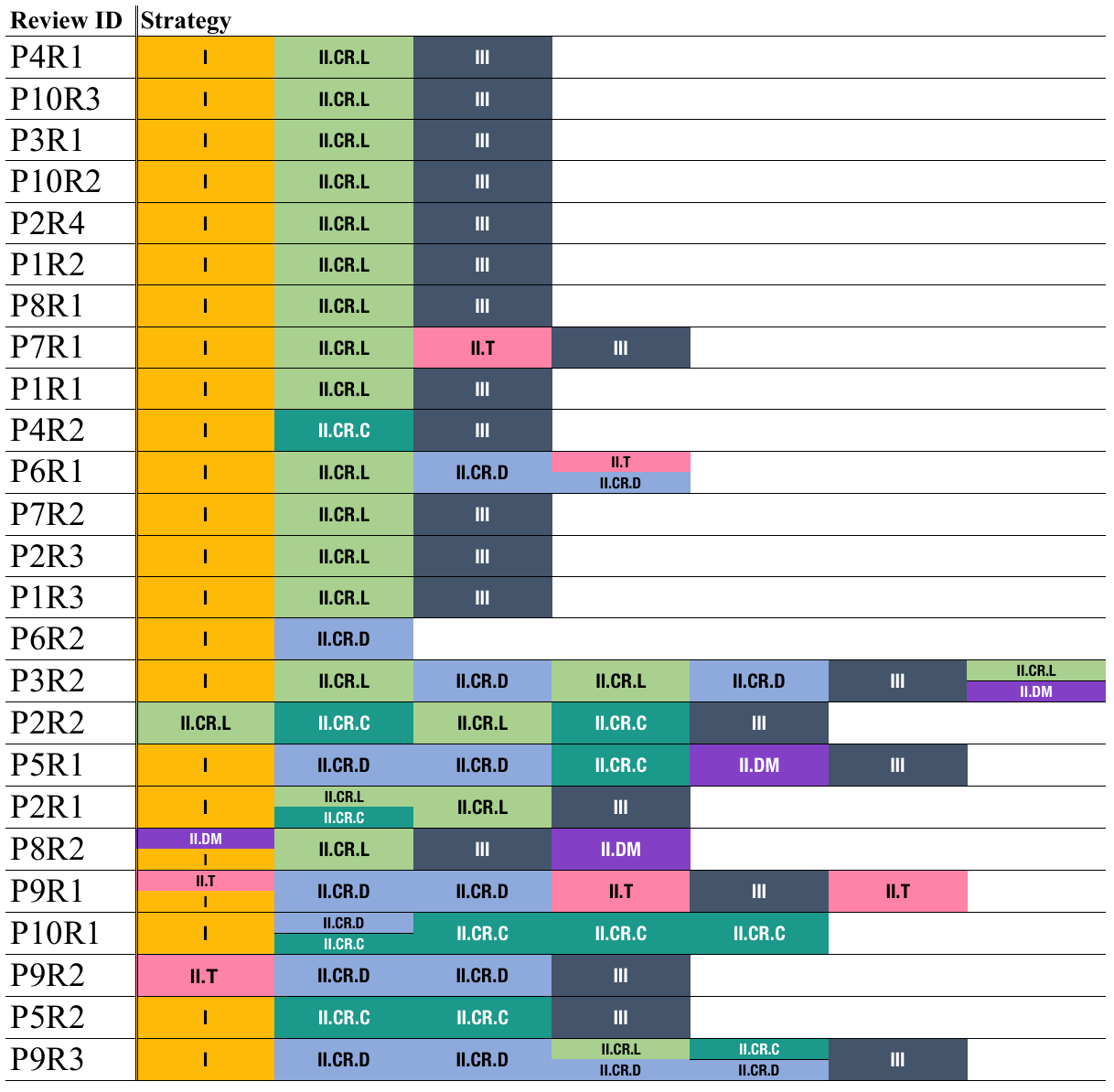}}
	\caption{Code Review Process Strategies}
	\label{fig:process}
\end{figure}

Understanding in code review is reached through multiple activities, presented in \Cref{fig:process} (a) --- with activities numbered and color-coded to guide the reader through the results.
\Cref{fig:process} (b) uses this color scheme to represent the sequence of activities across review sessions, ordered by change size.

In most reviews (N=23), the reviewer began with a \circled{I}~\emph{context building} phase, then proceeded to \circled{II} \emph{code inspection}, and concluded by submitting feedback, making a \circled{III} \textit{decision}, and potentially merging the change.
During the code inspection, reviewers  \circled{II.DM}~\textit{manage discussions}, perform \circled{II.CR}~\textit{code reading} using various strategies, and perform \circled{II.T}~\textit{testing}. While reading code, we observed that reviewers tended to follow a \circled{II.CR.L}~\textit{linear} reading approach for smaller changes (up to P1R3 - 6 files changed, 176 lines of code); while, for larger changes, they employed alternative strategies, such as \circled{II.CR.D}~\textit{difficulty-based} reading or \circled{II.CR.C}~\textit{chunking}, which enabled them to split the review into manageable units.

\smallskip
\noindent\circled{I}~\tagI. This step starts with reviewers familiarizing themselves with the PR title and description, and gathering other information sources (mentioned there or in the PR history and discussion; see \Cref{sec:information_sources} for more details). This initial step gives them an idea about the completeness of the available information and the Git hygiene~\cite{git_hygiene} of the PR.
It helps reviewers identify issues needing attention, actions required from them, and open questions.
The context-building phase also enables reviewers to build a preliminary mental model of the PR and assess its complexity.
Furthermore, reviewers form expectations about the PR and its ideal implementation, which they compare with what they see during the code inspection.
In the team of P10, the PR descriptions can also explicitly include the PR acceptance criteria or suggestions for the reviewing strategy. For instance, P10R1 included a suggestion to review the PR commit by commit, which P10 followed: \pquote{I would at least give it a try without even thinking about it.}
Yet, the benefits of this phase are compromised when the information sources are not provided.

Reviewers reduced the need for context building during the review session by being part of creating issues before the review or holding planning and alignment meetings with their team and the change's author. In P2R2 and P9R2 the context-building phase was skipped entirely due to a high availability of the context. Five reviewers (P2, P5, P7, P8, P9, and P10) underlined the importance of having a pre-alignment with the PR's author for review efficiency. As P10 explained: \pquote{We already pre-agreed on solutions so I don't need to challenge so much. It's more like does this match my expectation of what would happen or not? So [the review] is much smoother.}

\smallskip 
\noindent\circled{II} \tagII. This phase, shown in \Cref{fig:process}(a), consists of discussion management, code reading, and testing: 

\circled{II.DM}~\tagIIDM. \remove{Reviewers manage discussions at the beginning or end of a review, where reviewers engage in the discussions (P3R2, P5R1, P8R2).}\rev{While reviewers mostly wrote comments while performing Code Reading, some of the reviewers allotted a specific slot in their review to read through the available discussion and interact with the comments of other reviewers.}
This step was used to see what other reviewers and the author have already addressed, identify areas of agreement, and identify where they might add a new perspective. As P5 put it: \pquote{maybe the discussion already explains to me that I shouldn't review it right now, because somebody already went deep into this, and my time is not even needed right now here.}

\smallskip
\circled{II.CR}~\tagIICR. The strategies discussed here are inspired by \citet{heinonen2023synthesizing}. We observed three approaches: \emph{linear} reading, \emph{difficulty-based} reading, and \emph{chunking}.
%

~~\circled{II.CR.L}~\tagIICRL. Reviewers use this strategy with reviews of manageable size. 
As P2 puts it: \pquote{There are not many things that can go wrong in one line of code}. 
Linear reading was the main strategy used in small reviews until the review of P1R3 (176 changed LOC), as shown in \Cref{fig:process} (b).

~\circled{II.CR.D}~\tagIICRD. Reviewers employed this strategy when review size increased and they had to deal with the growing complexity. Reviewers prioritize what to review based on reviewing difficulty. \Cref{tab:difficulty} summarizes examples of code that reviewers deemed difficult or easy to review. 
Reviewers also referred to these two categories as areas where they (do not) need to invest significant effort and energy. The difficulty-based reading appears in two forms: (1) the \emph{easy-first} approach (P3R2, P6R1, P9R1, P9R2, P9R3) where reviewers first ``get rid of easy stuff'' then focus on the parts that require more effort, and (2) the \emph{core-based} approach (P5R1, P6R2) where reviewers began with what some of them refer to as ``the core of the change,'' then follow the data and execution flow to understand how the core changes were used in the code.

~\circled{II.CR.C}~\tagIICRC. This approach also deals with increasing complexity and can exist in parallel to top-bottom or difficulty-based reviewing. 
Reviewers may review only selected parts of a change, leaving the rest for later, or break the PR into chunks to narrow their scope and reduce their cognitive load. However, reviewing chunk by chunk can lead to a lack of overview; for example, to compensate for this drawback, the reviewers in P2R2 and P5R2 finalized their review by linearly reading the entire change to ensure the change's consistency.
Reviewers employed various units of chunking (listed in \Cref{fig:process}). Those reviewing the PR commit-by-commit stressed the importance of good Git hygiene and team processes to enable an effective breakdown of the review via commits. Other units, such as functional areas -- \eg models, migrations, and tests (P8R2), also served as ways to mentally segment the PR. Tests, in particular, were a significant unit of chunking. Reviewers reported different preferences as to when to review tests. P2 and P5 sometimes like to start with tests as they document the intention of the author, similarly to \emph{test-driven review}~\cite{spadini2019test}. In contrast, P8 preferred reviewing tests last.
Re-reviews of the same PR present a specific scenario where reviewers use \emph{comment-based} chunking (P2R1, P2R2, P5R2) to selectively check how were their previous comments addressed by the author.
Furthermore, we observed reviewers chunking their review based on \emph{files, issues,} \etc
%

\begin{table}
\caption{Examples of code that reviewers prioritized according to reviewing difficulty}
\label{tab:difficulty}
\begin{tabular}{p{3.5cm}p{4.5cm}}
\hline
\textbf{Hard to review}                           & \textbf{Examples}                                                                                                                       \\ \hline
Large changes                                     &                                                                                                                                         \\
Complex changes                                   & Logic, chained events                                                                                                                   \\
Potential for substantial issues                  & Parallel processing                                                                                                                     \\ \hline
\textbf{Easy to review}                           & \textbf{Examples}                                                                                                                       \\ \hline
Not author-written code                           & Auto-generated files, boilerplate code, binary files                                                                                    \\
Not production critical                           & Documentation, tests                                                                                                                    \\
Structurally constant changes                     & Declarative changes, established patterns in the code                                                                                   \\Small changes                                     & Renaming of program entities, small logic changes, string changes, upgrades, white space changes, usage of the main implemented element \\
Changes with low effect on the final code quality & Unnecessary removed files, visual front-end changes, code that will change in the future                                                \\ \hline
\end{tabular}
\end{table}


\circled{II.T}~\tagIIT. During code inspection reviewers not only review tests but also actively \emph{test} the PR (P6R1, P7R1, P9R1 and P9R2). They check the expected and actual output of functions in their local terminal, verify whether the system runs properly in their local environment with the new changes, perform hands-on testing of the responsiveness of the implemented UI changes, or troubleshoot failed CI/CD checks.

\smallskip
\noindent\circled{III}~\tagIII. Reviewers submit their comments, give overall feedback, and provide a verdict: accept the change, leave comments, or request further changes.
They finalize the review once they have checked all necessary parts of the PR, reached a desired level of understanding, provided sufficient feedback, know that they have no pending notes, and wrapped up any ongoing discussions. 

\subsection{\textbf{RQ3}: Role of information sources, knowledge base, and mental models}

Code review comprehension strategies rely on the information sources that include the reviewed code change and the reviewer's knowledge to construct a mental model of the code change. The mental models are then used to update the knowledge base, compare with other mental models (such as mental models of expected and ideal solutions), update them, and consequently evaluate the code change and provide feedback. In the following section, we provide an overview of the role of these CRCM components in review comprehension.

\subsubsection{Information Sources}
\label{sec:information_sources}

Reviewers use many information sources during code review. The ones explicitly mentioned in the most reviews are the PR title and description (21 reviews out of 25), the issue tracking (11 reviews), and the PR discussion (10 reviews). Reviewers use (1) information sources linked within the PR itself (\eg review size, commit titles, or CI/CD status), (2) resources to understand the broader code context of the PR (\eg code base, tests, or documentation), (3) tools to evaluate and test the change (\eg local development environment, specialized software tools), and (4) external sources not directly connected to the software system (\eg language documentation, ChatGPT, blog posts).
P7 and P9 both used specialized code review applications developed within their companies to \emph{test} the system behavior after a patch is applied and navigate the code base history.

We classified in which review stage reviewers used different resources. During the \emph{context-building} phase, which occurs at the start of the review, most of the information, especially what is presented and linked within the PR, is gathered. Interestingly, in 20\% of the observed sessions, reviewers needed to refer to other PRs for more context.
While PR discussion is accessed throughout the review, specific activities rely on different information sources. For instance, in the code inspection phase, reviewers use resources that help them navigate the code base, evaluate whether the PR was implemented correctly, and test the PR. Reviewers tested the PR in their local instance of the software system and used other specialized tools. Before making the final verdict, some reviewers revisited information sources such as CI/CD status or to-do notes.

\subsubsection{Knowledge Base}
\label{sec:knowledge_base}

This component of the model plays a key role in code review comprehension and in directly evaluating the PR for potential issues. Letovsky's model of code comprehension provides a broad depiction of knowledge that understanders use: from high-level domain knowledge and programming plans to the knowledge of programming language semantics. These fundamental aspects of code comprehension enable reviewers to recognize the `correct' implementation of the domain concepts, engineering processes, programming goals, and code formatting, and to set an expectation for code quality~\cite{SLMT24}. We observed that reviewers also employ their knowledge gained through their own experience as software developers, acquired through their interactions with code and other developers. This knowledge ranges from technical experience and familiarity with tools to their strategies to solve programming problems, to knowledge of their colleagues' expertise, coding and working style, and learning needs. Reviewers use this knowledge to provide more focused feedback: \pquote{If you don't have specific [coding] rules because you are a junior developer I will help you create them} (P6).

The more familiar a reviewer is with the overall context of the PR, the fewer resources they need to understand the PR and related artifacts. This helped enhance their ability to give constructive feedback for improvements. As mentioned in the \emph{context-building} phase, activities such as issue writing or pre-alignment with the team before the review reduce the need to build the context from information sources within the review itself. For instance, P1 quickly navigated code and provided feedback without stopping to understand or think further. When asked about it, they explained that it was due to their role in gatekeeping all code changes in the project and their extensive knowledge of the code base. They further clarified: \pquote{I have been reviewing code for a while, but there will be always changes that are completely new. And in those cases, it will take me longer. It could take even an hour to go through the test, match it with what I know, and figure it out}. The knowledge base also contains \emph{efficiency knowledge}~\cite{letovsky1987cognitive}, which refers to the explicit knowledge of common issues that can be directly applied to identify them.
This shows the fundamental role the knowledge base plays in context building and improving the efficiency and thoroughness of code reviews.

Below we report on the mental models reviewers construct during their review sessions. The knowledge base may already contain mental models \rev{stored in long-term memory that were constructed prior to the review and that} can be used during a review session: (1) a mental model of the software system---keeping knowledge of the processes, standards, expectations, and coding patterns in the code base that allows them to identify inconsistency with the system architecture, company and team coding practices; and (2) a mental model of the PR---reviewers may already have enough context about the PR to form a preliminary mental model of it.

The mental models of the system and the PR are constructed and updated through review iterations and stored in long-term memory for future recall. Thus, there is an ongoing exchange between the mental models in the knowledge base and the ones resulting from the comprehension process in the review session itself.

\subsubsection{Mental Models}
\label{sec:mental_model}

A mental model is the developer's mental representation of a program or code entity~\cite{heinonen2023synthesizing}---the PR in code review context.
In Letovsky's representation~\cite{letovsky1987cognitive}, the mental model is constructed in three layers: the specification, the implementation, and the annotation layer.
Here, we present in detail (1) the three layers of the mental model of the PR and (2) the use of alternative mental models of the PR as expectations and ideals that can be used for comprehension as well as PR evaluation.

\noindent\textbf{Layers of the Mental Model of the PR:}
An essential part of constructing the mental model of the PR is establishing the modification and increment to the software system---what parts of the system were changed and the size and complexity of these changes. Reviewers achieve this by using the available information sources (\Cref{sec:information_sources}) and by understanding the code itself (\eg by comparing the removed and added parts of the code, code tracing, or observing the system's behavior with the changes applied). This overview aids them in populating all three layers of their mental model.

\emph{Specification} refers to an explicit, complete description of the program's goals~\cite{letovsky1987cognitive}. Reviewers need to understand the review goal, \ie the problem being solved, the reasons for the change, and the scope of addressed cases. As they construct the specifications of the mental model of the PR, they also set explicit expectations for what the implementation should include, such as which tasks and code units the PR may contain, thus defining the evaluation criteria for the implementation. For instance, in the review \emph{P5R1}, the reviewer reads the title of a PR that introduces a counter for lazy loads on a specific object. The reviewer anticipates that the author probably counts lazy loads per request and logs them, which aligns with the actual implementation found later in the review.

The \emph{implementation} layer of the mental model refers to the actions and data structures in the program~\cite{letovsky1987cognitive}. Reviewers need to understand what has changed in the implementation and use code tracing~\cite{qi2020unlimited} to infer the code's behavior and its output. They also need to assess the rationale behind implementation choices and whether the changes are located correctly in the code.
Reviewers in the first review iterations on the PR might focus on ensuring the overall logic is sound before they dive into the implementation details. For instance, P10 explained: \pquote{I was thinking if it needs some other mapped table ... but that doesn’t matter that much as it is a technical thing and I am now checking mostly whether it makes sense}.

The \emph{annotation} layer connects specification goals to the parts of the implementation that fulfill them and which parts of the implementation fulfill certain specifications~\cite{letovsky1987cognitive}. As suggested, the annotation is done by reviewers by 
using top-down and bottom-up processes employed in parallel. When reviewers build their mental model in the context-building phase of the review, they can already construct all three layers. Many information sources allow them to build specifications and create expectations that are merely confirmed in a top-down manner during code inspection. The top-down annotation is supported by meaningful traceability across the issue tracking, PR title, and description, commit messages, file and variable naming to the implementation layer or by tracing review comments to their fixes. The bottom-up annotation starts by understanding code behavior and then assigning purpose to it. Reviewers commonly comprehended a piece of the implementation and noted to themselves that it `made sense'(P2, P3, P7, P8, P9, P10).

These three layers of the mental model create a complete understanding. Therefore, the review process can be streamlined by supplying reviewers early with cues that help them identify the change's goals and annotate them to expectations on how these goals are implemented. The expected version of the PR might, however, prove to be different from the actual PR implementation and specification during the review.

\noindent\textbf{The Expected and the Ideal:}
Reviewers form alternatives, variants, and extensions of their mental model of the PR in the form of expected and ideal solutions.

As mentioned in Sections \ref{sec:knowledge_base} and \ref{sec:mental_model}, reviewers tend to form expectations about the PR, which can reach the details of a partially formed mental model. We call this mental model the \emph{expected} mental model. Such a pre-existing model only needs to be confirmed when reviewing the PR. These expectations can stem from various sources, such as previous review rounds and entered comments, pre-alignment on solutions with the author, or general software engineering practices. Adhering to the expectations is desirable in the eyes of the reviewer (P8: \pquote{This is pretty much what I expected to see. So I'll just approve it}.) and makes the review more efficient (P10: \pquote{We already pre-agreed on solutions. So ... it's more like: Does this match my expectation of what would happen or not?})

An alternative set of models are the \emph{ideal} mental models, representing the optimal, corrected, or improved version of the PR. P7 remarked: \pquote{Now I'm wondering what the ideal solution is...}. These models are supported by the reviewer's programming experience, personal preferences, and knowledge of good engineering practices. Adherence to this imagined optimum is valued but not necessarily followed. P10 puts it: \pquote{In the end I don't necessarily agree that this is a better solution, but it still somehow solves the problem and it doesn't add a technical debt or anything like that}.

Both the expected and the ideal mental models offer an alternative or extension to the mental model of the actual PR. They may not always be formed or used, but when they are, they can be a vital tool to perform the review. Overall, mental models help reviewers set their expectations and evaluate changed codes against them, thus streamlining the review and identifying areas of potential improvements.

\section{Discussion}
\label{sec:discussion}

In this study, we observed 25 review sessions to analyze reviewing strategies through the lens of code comprehension.
Extending Letovsky's model of code comprehension~\cite{letovsky1987cognitive}, we developed the Code Review Comprehension Model, detailing how its components -- the information sources, knowledge base, and mental models interact and shape code review.



\subsection{Findings}

\noindent\textbf{Code comprehension is central to code review.}
Reviewers' ability to perform code review is dependent on their ability to understand the code change~\cite{wurzel2023competencies, bacchelli2013expectations, fagan1999design, macleod2017code}.
Letovsky's model~\cite{letovsky1987cognitive} offers a valuable framework to explain how comprehension shapes review, because reaching an understanding is among the reviewers' criteria to finalize their review and accept changes and drives them to seek information sources and choose effective reviewing strategies. \rev{Using the constructivist perspective on code comprehension~\cite{rajlich2002role}, we described how reviewers update their knowledge of programming practices and the software system through reviewing code. By viewing code comprehension as a tool to perform software development tasks, such as in the work of \citeauthor{von1993program}, we also described how the comprehension process contributes to evaluating the code changes and providing feedback.}

\smallskip
\noindent\textbf{Code review comprehension is scoped.}
According to \citet{letovsky1987cognitive}, full understanding is achieved when all three layers of a mental model are fully developed. In practice, full understanding can be reached over the entire PR lifetime, rather than in a single review session. \rev{As predicted by Piaget's theory of cognitive development~\cite{so1964cognitive}, we observed that developers scope their reviewing and understanding.} Full comprehension may not always be the goal, especially when reviewers perform focused, partial, or shallow reviews.
Scoping down the review is a commonly used strategy to deal with review complexity---one of the main challenges reviewers face~\cite{kononenko2016code}.

\smallskip
\noindent\textbf{Code review comprehension is incremental, iterative, and interactive.}
Reviewers already have mental models of the software system and may already have one for the PR itself before starting their review sessions. This knowledge allows them to be efficient when reviewing the PR.
Reviewers' mental model of the PR develops over time through review iterations and is used to update their mental model of the software system. Importantly, reviewers interact with the author and other reviewers to reach the desired level of understanding.

\smallskip
\noindent\textbf{Reviewing is strategic.}
Ad hoc reviewing is considered unsystematic~\cite{ciolkowski2003software, uwano2006analyzing}. However, we have observed that reviewers employ opportunistic strategies, similar to code comprehension, to reach an understanding of the PR and review the change.
Reviewers combine several activities in a modular way to form their reviewing strategy. First, they build the context. Then, they combine code reading, discussion management, and testing to comprehend and evaluate the change. The scoping and the modular design of their reviewing strategy allow them to creatively deal with change complexity.

\smallskip
\noindent\textbf{Ideals and expectations.}
Reviewers approach the review with ideas on ideal and optimal solutions and other expectations towards the PR, informed by common engineering solutions, good practices, and their knowledge of the PR as well as the software system. \rev{Using the self-discrepancy theory~\cite{bak2014self}, we could interpret the data effectively, identifying developers alternative mental models of the ideal and expected code changes.}
Meeting these ideals and expectations is seen as correct and streamlines the review by aiding reviewers to efficiently interpret the code, give feedback, suggest alternatives, and identify issues.

\subsection{Implications and Recommendations}

Based on these findings, we discuss recommendations on how to support reviewers in effective code review comprehension and change evaluation.

\smallskip
\noindent\textbf{For Code Review Tools:}
Our results provide insights for designing code review tools that better align with reviewers' natural strategies. First, reviewers scope their review sessions and prioritize their activities. Tools can support reviewers in viewing only certain aspects and parts of the PR, as well as provide ways to let reviewers communicate the scope and main outcomes of the performed review.

The main strategies for code reading while dealing with review complexity were chunking and difficulty-based reading. This finding is an opportunity to investigate how tools can provide support to identify and visually separate meaningful code chunks, indicate the change distribution by signifying the core of the change or more complex passages, and support Git hygiene.
Reviewers often switched contexts to navigate the code base or test the code change. Both these activities were substituted in two cases (P7, P9) by an in-house reviewing application. This behavior shows there is a need to integrate code base navigation, testing, and reviewing in one environment together with information sources, such as issue trackers or online/documentation search.

Reviewers used review comments as an information source throughout the review, while remaining wary that comments could be a source of potential bias. Tools should be designed to allow reviewers to be able to toggle the viewing of existing review comments.
Furthermore, reviewers created many mental models of the code change, their expectations, and ideals. Tools can support developers to collaboratively construct and cross-reference multiple mental models~\cite{storey1999cognitive}.

Future research can be designed and carried out to investigate whether and how AI and LLMs have the potential to support many of the aforementioned improvements, \eg by suggesting effective reviewing strategies, annotating review comments to related fixes in the implementation, or generating missing context and documentation~\cite{sakib2024automatic}.

\smallskip
\noindent\textbf{For Practitioners:}
We observed real-world code reviews performed by experienced reviewers recommended by others as great reviewers---their practices can be used to propose effective reviewing strategies.

For \textit{dealing with review complexity}, our participants mainly used three strategies: (1) narrowing down the scope of the review, (2) using iterations to reach full understanding one step at a time, and (3) adjusting the reading approach (\eg using chunking). The effectiveness of the first strategy is reinforced by previous research, which found that focused reviews enhance reviewers’ ability to detect defects~\cite{braz2022less}.
To better support the third strategy, good Git hygiene is helpful; and, if there is a clear core of the change, it can be used as a starting point for reading.

For \textit{improving reviewing efficiency}, our participants relied on a strong knowledge base, by knowing programming practices, the codebase, and the team. Additionally, they also relied on the colleagues involved in the review process to complement their expertise and understanding. Finally, creating expectations about the change under review and having an alignment with the change author prior to the review session was used to improve reviewing effectiveness.

\smallskip
\noindent\textbf{For Researchers:}
Code comprehension is a stepping stone in performing software development tasks~\cite{von1995program}. \citeauthor{letovsky1987cognitive}'s model of code comprehension has proved to be good basis to start describing how code review is performed in a real scenario, including the diversity of reviewers' strategies and information sources. Using this model can be applied to other software development tasks to formulate recommendations towards more human-centric tool design. 

Reviewers' activity has been previously approached with methods such as experiments~\cite{gonccalves2022explicit, fregnan2022first, dunsmore2000role}, interviews~\cite{spadini2019test}, eye-tracking~\cite{sharif2012eye}, and mining software repositories~\cite{braz2022less, pascarella2018information, fregnan2022first}.
However, these methods have limitations in capturing critical factors such as context, priorities, interactions among reviewers, and their knowledge of the software system being reviewed.
Usually, eye-tracking studies and experiments use artificial code changes of a manageable size where linear reviewing strategies were mostly observed~\cite{sharif2012eye, fregnan2022first}.
However, we have observed that complex reviews in real-world context require strategies to manage the complexity. These strategies include narrowing the scope of the review or concluding it when the reviewers feel their comments no longer provide significant value, or when they run out of time or energy to continue the review.
This behavior may explain why \citet{fregnan2022first} observed a linear decrease of the number of comments in each subsequent file of the PR or contextualize the test-driven review among other reviewing practices~\cite{spadini2019test}.

\smallskip
\noindent\textbf{For Educators:}
Students benefit from being taught (1) explicit programming strategies and (2) to reflect on their practices~\cite{ko2019teaching}. Our findings can aid students to reflect on their reviewing strategies. Moreover, students can be instructed on potential reviewing strategies, decomposing PRs in more reviewable units, and finding appropriate information resources.
Students can learn to leverage the experience and expertise of other developers and create rich expectations for change evaluation through building rich context and documentation of PRs and creating alignment with their teammates.
Educators should emphasize the importance of a strong foundation in programming patterns to enhance the effectiveness and efficiency of code reviews. This is particularly crucial as the use of AI-generated code becomes more prevalent, requiring students to develop the skills needed to review such code effectively.

\journal{

\appendix
\label{sec:Guidelines}

\textbf{Guidelines for Reviewers}

\emph{Before the review session}

\begin{itemize}
    \item Know your code base, know your tools
    \item Get involved in the discussion around the PR
    \item Get involved in writing issues
    \item Align with the author on vision around the PR - what are you going to achieve, what is the direction for the implementation
    \item Prepare your reviewing environment
\end{itemize}

\medskip

\emph{During the review session}

\begin{itemize}
    \item Gather context for the review from the available information.
    \item If you can keep all the code in your memory easily, proceed with the top to bottom review.
    \item When the changes are more complex, you can make a partial or focused review.
    \item Find out if commits or file structure allows for useful chunking of the review into meaningful units.
    \item Try to split your review in managable steps.
    \item Investigate the distribution of the change - see what was changed and where are the most core changes of the PR.
    \item Consider the added value of your review and feedback. If the added value is low, consider whether further feedback is needed, terminate your review or get in touch with the author for an in person meeting.
    \item Find out what colleagues can compensate in the review for areas you understand less.
    \item Indicate to other review participants what scope did your review cover and what are the main issues to be addressed in order to accept the change.
    \item When stagnating in yout review, consider finding new information resources or asking the author.
    \item Consider going through the PR once more to make a consistency check of the whole.
\end{itemize}

\medskip

\emph{Identifying issues} 

\begin{itemize}
    \item Stuff
\end{itemize}

\medskip

\emph{Concluding the review}

\begin{itemize}
    \item Concluding
    \item Giving verdict
    \item Writing the review message
\end{itemize}

\medskip

\textbf{Guidelines for the author}

\begin{itemize}
    \item PR description
    \item link all that might be relevant
    \item suggest strategy
    \item evaluation criteria, definition of done
    \item Git hygiene
    \item Alignment
    \item Indicate to reviewers how were their comments addressed
    \item Expectations - specification, implementation, annotation
\end{itemize}

}

\section{Conclusion}
\label{sec:conclusion}
In the study presented in this paper, by observing expert reviewers---recommended as great reviewers by others---while reviewing real-world code changes and interviewing them afterwards, we could uncover, for the first time, the opportunistic strategies they use to form an understanding of the code change and to evaluate it.
Based on these strategies, we developed a Code Review Comprehension Model and put forward recommendations on how to improve code review tools as well as on topics to conduct further research.

\section*{Acknowledgment}

Alberto Bacchelli and Pooja Rani gratefully acknowledge the support of the Swiss National Science Foundation through the SNF Projects 200021\_197227 and 200021M\_205146. Pavl\'{i}na Wurzel Gonçalves gratefully acknowledges the
support of CHOOSE, the Swiss Group for Original and Outside-the-box Software Engineering (https://choose.swissinformatics.org/).


\bibliographystyle{plainnat}
\bibliography{cr_process}

\end{document}